\documentclass[12pt]{iopart}

\usepackage{graphicx}

\begin{document}

\title{Spatiotemporal dynamics of biocrust and vegetation on sand dunes}

\author{Hezi Yizhaq\ddag~and Yosef Ashkenazy\ddag
\footnote[3]{To whom correspondence should be addressed (yiyeh@bgu.ac.il)}}

\address{\ddag\ Solar Energy and Environmental Physics, BIDR, Ben-Gurion University, Midreshet Ben-Gurion, Israel}

\begin{abstract}
  We propose a model to study the spatiotemporal dynamics of biocrust and vegetation cover on sand dunes. The model consists of two coupled partial nonlinear differential equations and includes diffusion and advection terms for modeling the dispersal of vegetation and biocrust and the effect of wind on them. In the absence of spatial variability, the model exhibits self-sustained relaxation oscillations and regimes of bistability--the first state is dominated by biocrust and the second by vegetation. We concentrate on the one-dimensional dynamics of the model and show that the front that connects these two states propagates mainly due to the wind advection.
  In the oscillatory regime, the front propagation is complex. For low wind DP (drift potential) values, a series of spatially oscillatory domains develops as the front advances downwind. These domains form due to the oscillations of the spatially homogeneous states away from the front. However, for higher DP values, the dynamics is much more complex, becoming very sensitive to the initial conditions and exhibiting an irregular spatial pattern as small domains are created and annihilated during the front advance. Such irregular dynamics can be associated with the temporal variations of dune cover. In addition, similar behavior can be generated by other models that exhibit temporal oscillations and bistability.
\end{abstract}


\submitto{\NJP}

\maketitle

\section{Introduction}\label{sec:sec1}
Sand dunes cover up to one third of the arid regions of the low and mid-latitudes and form important, unique landscapes and ecosystems ~\cite{Lancaster-1995:geomorphology,Tsoar-2013:critical}. Dunes are dynamic bedforms and are sensitive to climate variability on a variety of spatiotemporal scales ~\cite{Thomas-Knight-Wiggs-2005:remobilization,Ashkenazy-Yizhaq-Tsoar-2012:sand,Yizhaq-Ashkenazy-Levin-Tsoar-2013:spatiotemporal}. Dunes are also a potential source of dust caused by aeolian abrasion emission ~\cite{Enzel-Amit-Crouvi-et-al-2010:abrasion}--dust has an important impact on loess formation~\cite{Crouvi-Schepanski-Amit-et-al-2012:multiple} and on marine ecological systems ~\cite{Bhattachan-Odorico-Baddock-Zobeck-Okin-Cassar-2012:southern}. In many places around the world, dunes are considered to be a threat since they affect human activity and property, such as roads, railroads, houses, and more. Substantial efforts have been made to stabilize dunes, such as in the green belt of China ~\cite{Dong-Chen-He-et-al-2004:controlling,Khalaf-Al-Ajmi-1993:aeolian}. However, in other regions, efforts have been made to increase dune activity by vegetation and biocrust removal, in part to enrich the biodiversity of psammophilous (sand-loving) species ~\cite{Rubinstein-Groner-Yizhaq-et-al-2013:eco}.

Dunes can be active (mobile), semi-active, or fixed (stable), depending on wind, precipitation, human activities, and dune cover (sand, vegetation, and biocrust) \cite{Thomas-Wiggs-2008:aeolian,McKenna-Neuman-Maxwell-1999:wind,Argaman-Singer-Tsoar-2006:erodibility,Ashkenazy-Yizhaq-Tsoar-2012:sand,Tsoar-2013:critical,Kinast-Meron-Yizhaq-Ashkenazy-2013:biogenic}. Vegetation and biocrust play important roles in dune stabilization, and they are affected by winds and precipitation. On the one hand, when dunes are covered by vegetation above a certain critical value ~\cite{Ash-Wasson-1983:vegetation}, even very strong winds are masked by the vegetation, such that they do not reach the ground and, thus, do not result in sand erosion and dune mobility~\cite{Mayaud-Webb-2017:vegetation}. In this case, the flow is termed ``skimming flow,'' where wakes completely overlap, and the entire surface is sheltered from wind erosion ~\cite{Suter-Burri-Gromke-Leonard-et-al-2013:spatial}. On the other hand, when the dunes are bare, even relatively weak winds that are above the threshold velocity (about 6 m/s)~\cite{Fryberger-1979:dune} lead to sand transport. Winds also directly affect vegetation by exerting stress on it and by increasing evapotranspiration, a condition that further reduces vegetation cover, hence leading to enhanced dune activity.

Biocrust (also known as microbiotic crusts or biological soil crusts) can be found on vegetated sand dunes throughout the world (mostly on linear dune slopes and on interdune areas), including the vegetated linear dunes (VLDs) of Australia, the northern Negev Desert, the Thar Desert, and the Kalahari Desert ~\cite{Hesse-Simpson-2006:variable,Danin-1996:plants,Tsoar-2013:critical,Thomas-2013:aeolian}. VLDs are shorter than unvegetated linear dunes; their height ranges from a few meters up to dozens of meters~\cite{Tsoar-2013:critical}.
In arid and hyper-arid environments, the biocrust is mainly composed of a thin cyanobacterial layer ~\cite{Kidron-Zohar-2014:wind,Almog-Yair-2007:negative}, but in wetter regions, it can also contain lichens and mosses~\cite{Almog-Yair-2007:negative}. In the northwestern Negev Desert’s sand dunes, biocrust consists of filamentous cyanobacteria, which serve as traps for atmospheric dust, since they increase the surface roughness ~\cite{Veste-Breckle-Eggert-Littmann-2011:vegetation,Zaady-Katra-Yizhaq-et-al-2014:inferring}. In turn, the deposited dust can contribute to crust development through its physical presence and chemical reactivity~\cite{Viles-2008:understanding,Rozenstein-Zaady-Katra-et-al-2014:effect}. The cohesive biocrust is more resistant to wind erosion than bare sand as the filamentous cyanobacteria of the crust bind grain particles together~\cite{Kidron-Zohar-2014:wind}. Wind-tunnel experiments showed that this crust’s resistance to wind erosion depends on the biocrust type~\cite{Neuman-Maxwell-2002:temporal}. Due to its resistance to wind erosion and to prolonged droughts, biocrust plays a crucial role in sand dunes’ surface stability ~\cite{Belnap-Lange-2001:biological,Siegal-Tsoar-karnieli-2013:effects,Amir-Kinast-Tsoar-et-al-2014:effect}. Biocrust destruction by the trampling of grazing animals leads to a sharp increase in aeolian sand transport rates~\cite{Neuman-Maxwell-2002:temporal} as occurred in the sand dunes on the Egyptian side of the Israeli-Egyptian border, due to intensive grazing of Bedouin herds and clear cutting~\cite{Tsoar-2008:land} .

The mutual relations between vegetation and biocrust are complex and depend on the successional stage of the biocrust after its establishment. It is still debated whether the presence of biocrust enhances vegetation growth or diminishes it~\cite{Almog-Yair-2007:negative,Kidron-Zohar-2014:wind,Bel-Ashkenazy-2014:effects}, and whether this mostly depends on whether the biocrust is hydrophobic (positive effect) or hydrophilic (negative effect). On other soils (such as loess), biocrust enhances vegetation growth through the so-called ``source-sink'' effect, meaning that runoff on the crust flows to sink points where the vegetation is present and the infiltration rate is higher~\cite{Shachak-Lovett-1998:atmospheric}. This feedback is also known as ``infiltration feedback'' and leads to vegetation pattern formation in water-limited systems ~\cite{Meron-2015:nonlinear,Yizhaq-Stavi-Swet-et-al-2019:vegetation}.

The activity of sand dunes is sensitive to climatic conditions, mainly to wind power and precipitation~\cite{Tsoar-2005:sand,Hugenholtz-Wolfe-2005:biogeomorphic,Yizhaq-Ashkenazy-Tsoar-2009:sand,Ashkenazy-Yizhaq-Tsoar-2012:sand,Tsoar-2013:critical,Warren-2013:dunes}, although their relative importance is still an open question \cite{Bogle-Redsteer-Vogel-2015:field}. The activity of aeolian deposits can change on time scales of $10-10^5$ years ~\cite{Thomas-Wiggs-2008:aeolian}. The use of a luminescence (OSL) dating technique showed that the dunes in the Kalahari, Central Asia, Australia, and the Negev ~\cite{Maman-Blumberg-Tsoar-et-al-2011:central,Roskin-Porat-Tsoar-et-al-2011:age} have undergone periods of activity and stability since the Late Pleistocene ~\cite{Tsoar-2013:critical}. These periods of different activity types are usually attributed to increased windiness and/or decreased precipitation ~\cite{Roskin-Porat-Tsoar-et-al-2011:age}. Dune mobility can also vary on decadal and inter-annual time scales~\cite{Lancaster-2013:climate}, due to changes in sediment supply, availability, and mobility that are determined by regional and local climate conditions and by vegetation and biocrust cover.

In a previous study, we examined the homogeneous (space-independent) system of biocrust and vegetation on sand dunes modeled by two coupled nonlinear ordinary differential equations (ODEs) ~\cite{Kinast-Meron-Yizhaq-Ashkenazy-2013:biogenic,Yizhaq-Ashkenazy-2016:oscillations} and showed that it can exhibit self-sustained oscillations with different periods (from 300 years to more than 3000 years) and amplitudes, even under constant climatic conditions (precipitation and wind). We proposed that this internal oscillatory dynamics of vegetation and biocrust may interact with the external climatic conditions on various time scales, and may result in richer dynamics of both the dune and climate systems. As VLDs constitute the main dune type utilized in continental desert paleo-aridity studies ~\cite{Thomas-2013:aeolian,Tsoar-2013:critical,Warren-2013:dunes}, because they are more prone to vegetation and biocrust colonization, it is important to understand the possible various aspects of their complex dynamics.

In a previous study~\cite{Yizhaq-Ashkenazy-Levin-Tsoar-2013:spatiotemporal}, we proposed a spatial vegetation model (without biocrust) to model the spatiotemporal dynamics of vegetation cover on sand dunes and focused on the dynamics of transgressive dunes. The model predicted the growth of a transgressive dune parallel to the wind and its shrinking perpendicular to the wind, where, depending on geometry and size, a transgressive dune can initially grow and then eventually shrink. The larger the initial area, the slower its stabilization process. The model's predictions were validated by field observations from Fraser Island in Australia. The model we propose below is based on this model and also includes the effect of biocrust cover on sand dune dynamics.

\begin{figure}[!ht]
\begin{center}
\includegraphics[width=8in]{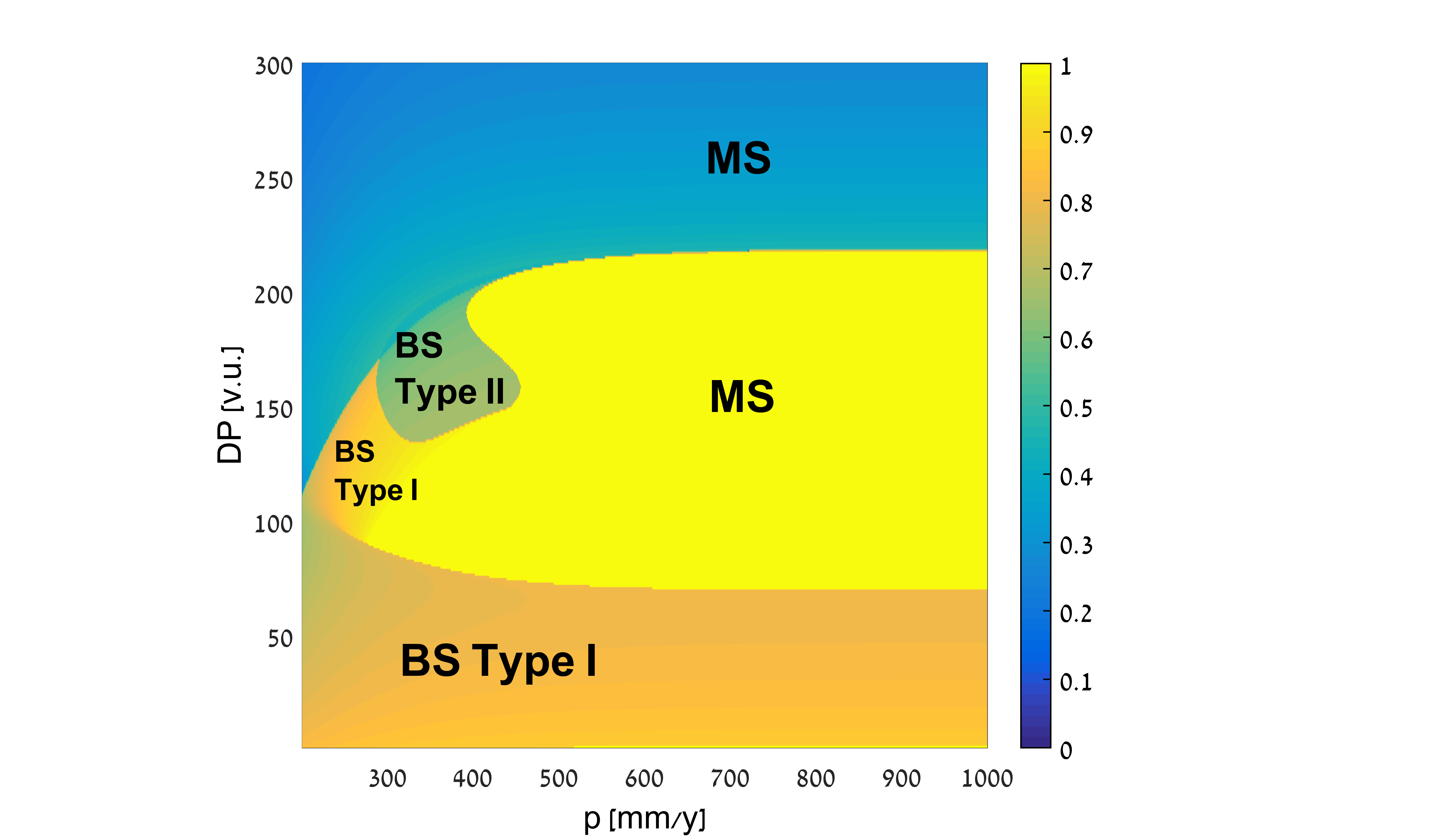}
\end{center}
\caption{Summary of the types of dynamics of the vegetation-biocrust space-independent model; this figure is basically Fig. 6 of \cite{Yizhaq-Ashkenazy-2016:oscillations}. The figure shows the different domains of stability and bistability for vegetation cover (homogeneous stable solutions); MS stands for mono-stable state, and BS for bistable states. BS Type I is a domain with two steady state solutions, whereas, in the BS Type II domain, the bistability is of one stationary solution and one oscillatory solution. The different solutions and the parameter values can be found in \protect\cite{Yizhaq-Ashkenazy-2016:oscillations}. $b$ exhibits a similar diagram. In this work, we study front solutions in BS Type I and BS Type II domains. Based on \protect\cite{Yizhaq-Ashkenazy-2016:oscillations}, the following set of parameters was used in most of the simulations: $\eta_v=0.2$, $\alpha_{max,v}=0.15$ $\rm yr^{-1}$, $\epsilon_v=0.001$ $\rm yr^{-1}VU^{-1}$ (VU are vector unit), $\gamma=8\cdot {10^{-4}}$ $\rm yr^{-1}VU^{3/2}$, $\mu_v=0.02$ $\rm yr^{-1}$, $\phi_v=0.01$ $\rm yr^{-1}$, $c_v=100$ mm/yr, $v_c=0.3$, $p_{min,v}=50$ mm/yr, $\eta_v=0.1$, $d=15$ for the equation for $v$, and $\eta_b=0.1$, $\alpha_{max,b}=0.015$ $\rm yr^{-1}$, $\epsilon_b=1\cdot {10^{-4}}$ $\rm yr^{-1}VU^{-1}$, $\mu_b=1\cdot {10^{-4}}$ $\rm yr^{-1}$, $\phi_b=0.01$ $\rm yr^{-1}$, $p_{min_b}=20$ mm/yr and $c_b=50$ mm/yr for the equation for $b$. This figure is modified from ~\cite{Yizhaq-Ashkenazy-2016:oscillations}. }
\label{fig:fig1}
\end{figure}

In the current work, we extend the homogeneous model to include spatial dynamics with terms representing vegetation and crust dispersal and the effects of the wind (direction and power) on dune cover when it is not homogeneous. We studied the one-dimensional front propagation/dynamics (i) when the front separates bistable states, i.e., one solution is dominated by vegetation and the other by crust, and (ii) when the dynamics in the oscillatory regime is such that each side of the front is in a different phase of the oscillation (Fig. \ref{fig:fig1}). In the latter case, the spatiotemporal dynamics is very complex, and it is hardly predictable for large DP (drift potential) values; all the model's parameters are temporally and spatially constant.

\section{The model}\label{sec:sec2}

The ODE (spatially independent) part of the model was introduced and studied in the context of the existence and stability ranges of different dune-cover states along gradients of rainfall and wind power ~\cite{Kinast-Meron-Yizhaq-Ashkenazy-2013:biogenic,Kinast-2014:spatial} and of the periodic relaxation oscillations in biocrust/vegetation dynamics ~\cite{Yizhaq-Ashkenazy-2016:oscillations}. We also introduced a spatial model for vegetation cover on sand dunes \cite{Yizhaq-Ashkenazy-Levin-Tsoar-2013:spatiotemporal} to study the spatiotemporal dynamics of transgressive dunes. Here we extend these two models and propose a model to study the spatiotemporal dynamics of both vegetation and biocrust on sand dunes.

The model consists of two coupled nonlinear PDEs (partial differential equations) for the vegetation cover $v$ and the biocrust cover $b$. More details on the ODE parts of the model can be found in~\cite{Kinast-Meron-Yizhaq-Ashkenazy-2013:biogenic,Yizhaq-Ashkenazy-2016:oscillations} and on the spatial terms in \cite{Yizhaq-Ashkenazy-Levin-Tsoar-2013:spatiotemporal}. The model's equations are
\begin{eqnarray}\label{eq:model}
  \frac{dv}{dt} &=& \alpha _v (v + \eta _v )s - (\varepsilon _v g + \gamma {\rm DP}^{2/3}  + \mu _v  + \phi _v b)v \nonumber \\
  && - {\beta _v}{\rm{DP}}\left| {\nabla g \cdot \vec k} \right|v - \kappa {\rm{D}}{{\rm{P}}^{2/3}}(\nabla v \cdot \vec k)v + {\delta _v}{\nabla ^2}v
  \\
  \frac{db}{dt} &=& \alpha _b (b + \eta _b )s - (\varepsilon _b g + \mu _b  + \phi _b v)b
  - {\beta _b}{\rm{DP}}\left| {\nabla g \cdot \vec k} \right|b + {\delta _b}{\nabla ^2}b \,
\end{eqnarray}
where $s\equiv 1-v-b$ is the bare sand cover; $v, b, s$ represent relative cover and are thus limited to be between 0 and 1. The first term on the LHS of both equations describes a logistic growth modulated by a growth rate parameter, which, for $p > p_{{\rm min},v,b}$, depends on precipitation $p$ as follows:
\begin{equation}\label{eq:alpha}
\alpha _{v,b}  = \alpha _{\max ,v,b} (1 - \exp ( - (p - p_{\min ,v,b} )/c_{v,b} )) \,
\end{equation}
and is 0 for $p \le p_{{\rm min},v,b}$ (taken as 20 mm/yr for the biocrust and 50 mm/yr for the vegetation); $\alpha _{{\rm max},v,b}$ is the maximum growth rate of vegetation and biocrust covers, respectively. The other terms on the LHS, which do not depend on space, are different mortality terms that are associated with the direct and indirect wind effect, natural mortality, and competition ~\cite{Kinast-Meron-Yizhaq-Ashkenazy-2013:biogenic,Kinast-2014:spatial}; we elaborate on these below.

Both biocrust and vegetation are suppressed by sand transport and sand blasting of saltating particles ~\cite{Okin-2013:linked}, which is modeled by the function $g$,
\begin{equation}\label{eq:g}
g = \frac{1}{2}{\rm DP}(\tanh (d(v_c  - v)) + 1)s \,
\end{equation}
where DP is the wind drift potential, which is a measure, in vector units, of the potential sand transport by the wind. It is derived from surface (10 m high) wind velocity above the threshold velocity for sand transport ($\approx$ 6 m/s), through weighting of the sand transport equation. Its definition is: ${\rm DP = }\langle U^2 (U - U_t )\rangle$, where $U$ is the wind speed (in knots: 1~knot~=~0.514~m/s) at 10~m height, and $U_t$ is the minimal threshold velocity (=~12~knots) necessary for sand transport~\cite{Fryberger-1979:dune}. There are both theoretical and empirical linear relations between DP and the rate of sand transport ~\cite{Bullard-1997:note}. The function $g$ is a step-like function and mimics the effect of the critical vegetation cover $v_c$ (i.e., $g=1$ for $v=0$ and $g\to 0$ for $v>v_c$).
The sand transport depends on the available bare sand $s$, and it becomes zero when the dune is fully covered by vegetation and biocrust ($v+b=1$). The other mortality terms ($-\phi _v b v$ and $-\phi _b b v$) represent the competition between the biocrust and the vegetation and mortality due to grazing and trampling ($-\mu _v v$ and $-\mu_b b$, respectively). In addition, there is a mortality term of vegetation, $-\gamma {\rm{DP}^{2/3}} v$, that accounts for vegetation decay due to direct wind action (this term exists even without sand transport), which increases evapotranspiration and stress, enhances branch breaking, and thus, limits vegetation growth~\cite{Hesp-2002:foredune}. The two-thirds power aims to represent the wind drag on vegetation, which is proportional to the square \cite{Bagnold-1941:physics} of the wind speed $U$ (while DP$\propto U^3$). $\gamma$ is a proportionality parameter that may depend on vegetation types as described in detail in \cite{Yizhaq-Ashkenazy-Levin-Tsoar-2013:spatiotemporal,Yizhaq-Ashkenazy-Tsoar-2009:sand}. Note that the equation for the biocrust does not include such a mortality term since biocrust can withstand very strong wind ~\cite{Neuman-Maxwell-2002:temporal}.

The last three terms in the equation for the vegetation cover, $v$, describe its spatial dynamics acting in regions where it is spatially variable, e.g., fronts that separate active and stabilized domains (i.e., areas where $v$ is not spatially uniform). These terms represent the effect of the intensity and wind direction on vegetation cover using: (i) a mortality term due to sand erosion or deposition, $- {\beta _v}{\rm{DP}}\left| {\nabla g \cdot \vec k} \right|v$, (ii) a term representing the direct suppression of the wind on vegetation, $- \kappa {\rm{D}}{{\rm{P}}^{2/3}}(\nabla v \cdot \vec k)v$, and (iii) a diffusion term representing the spatial spreading of the vegetation cover, ${\delta _v}{\nabla ^2}v$. More details about the derivation of these terms can be found in \cite{Yizhaq-Ashkenazy-Levin-Tsoar-2013:spatiotemporal,Yizhaq-Ashkenazy-Tsoar-2009:sand}.
The ``$\beta$'' term is zero where the vegetation cover is uniform, and it is negative at the border between active and stabilized dunes in both directions of the prevailing winds. In the current version of the model, we exclude the effect of pioneer species and psammophilous plants, which are tolerant to sand deposition and erosion~\cite{Nield-Baas-2008:influence,Bel-Ashkenazy-2014:effects}.

The last spatial term in the equation for $v$, $ \delta_v \nabla ^2 v$, represents an isotropic seed dispersal (e.g., by animals and winds) where $\delta_v$ is a diffusion coefficient. Although the wind can induce long-distance seed dispersal, most of the seeds fall at a short distance from the canopy ~\cite{Nathan-Katul-Horn-et-al-2002:mechanism}, which justifies the use of a simple diffusion term and not a more sophisticated kernel function.  It is important to note that under this model's assumptions, the diffusion term acts both parallel and perpendicular to the wind direction, whereas the former two spatial terms act only along the wind direction. The diffusion term acts to smooth the vegetation cover at the borders between bare and vegetated areas. Thus, in addition to its physical representation of vegetation growth, this term prevents the development of numerical instabilities due to unbounded vegetation gradients at the transition zones between bare and vegetated domains, thus helping to smoothen the fronts in the bistability regime.

The equation for the biocrust cover, $b$, includes two spatial terms. The first, $- {\beta _b}{\rm{DP}}\left| {\nabla g \cdot \vec k} \right|b$, describes the effect of sand transport on the biocrust when there is a gradient in the vegetation cover (note that $g$ does not depend on $b$). It is the analog term of the term $- {\beta _v}{\rm{DP}}\left| {\nabla g \cdot \vec k} \right|v$ in the equation for the vegetation cover. As for the parallel term in the vegetation cover equation, it is negative in both signs of the vegetation gradient since both sand erosion and sand deposition suppress biocrust growth and are mostly modulated by the vegetation cover. The second term, $ \delta_b \nabla ^2 v$, represents the spreading of biocrust due to a diffusion-like process (here we assume only short-range biocrust dispersal), where $\delta_b$ is a diffusion coefficient that is much smaller than $\delta_v$. Note that in the equation for the biocrust cover, there is no spatial term that models the direct effect of wind on biocrust since it can withstand very strong winds. More details about the model and its spatial version (which only accounts for vegetation cover) can be found in ~\cite{Yizhaq-Ashkenazy-2016:oscillations,Kinast-Meron-Yizhaq-Ashkenazy-2013:biogenic,Yizhaq-Ashkenazy-Levin-Tsoar-2013:spatiotemporal,Yizhaq-Ashkenazy-Tsoar-2009:sand}.

\section{Results}\label{sec:sec3}
\subsection{Front dynamics in the BS Type I domain}\label{subsec:sec1}

\begin{figure}[!ht]
\begin{center}
\includegraphics[width=0.8\columnwidth]{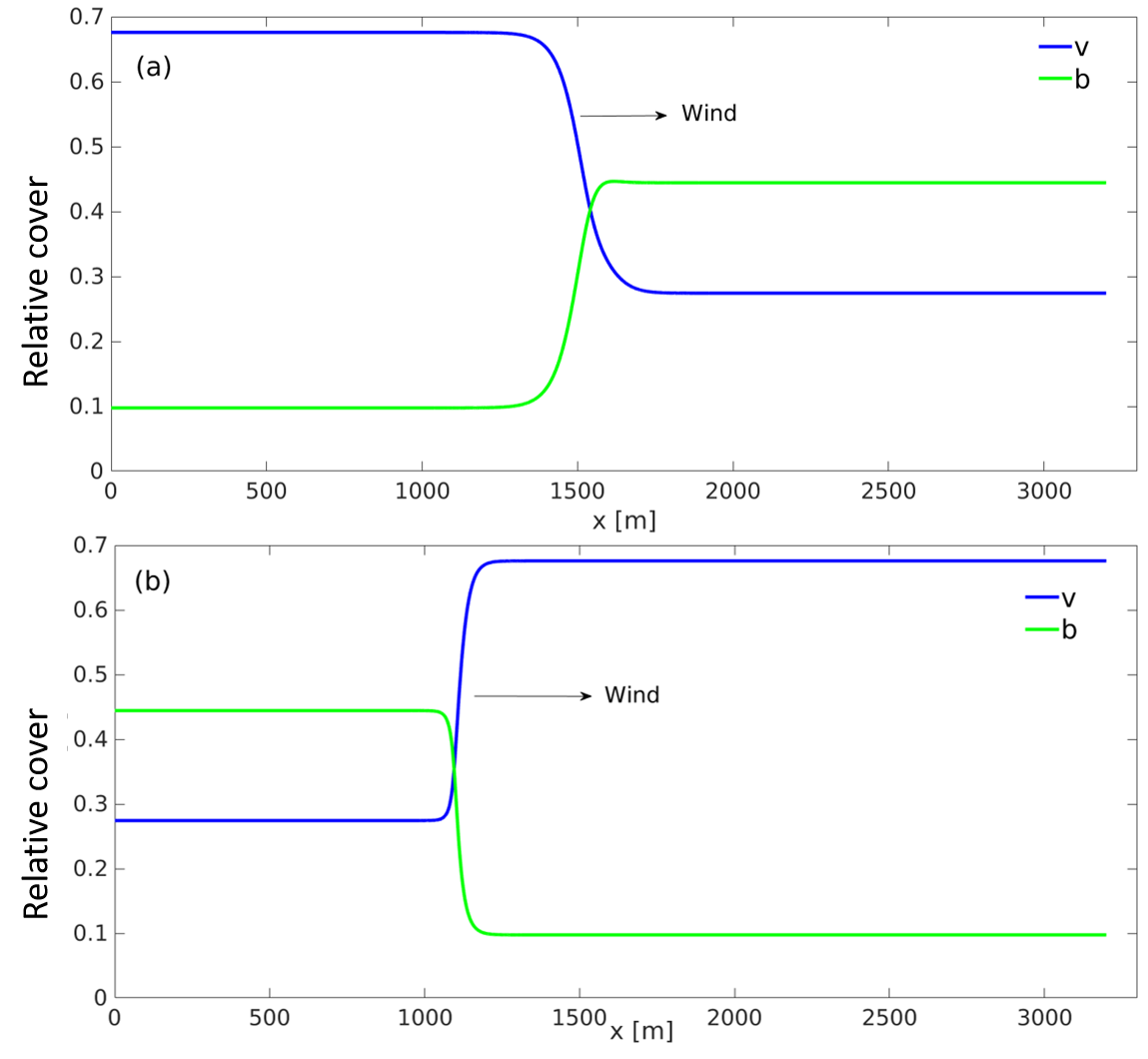}
\end{center}
\caption{1D front solutions for DP=120 and $p=300$ mm (BS Type I fronts in Fig. \ref{fig:fig1}). (a) $v_{up}$ front solution that connects a vegetation-dominant state with a crust-dominant state. (b) The ``reversed'' scenario of (a), i.e., $b_{up}$ front solution that connects a biocrust-dominant state with a vegetation-dominant state. The wind direction in both cases is from left to right as indicated by the arrow. Parameters values of the five spatial terms that have been used in all the simulations are:
$\beta_v=6.64\cdot10^{-4}$, $\kappa=0.1$, $\delta_v=0.1$, $\beta_v= 6.64\cdot10^{-5}$, $\delta_b=0.2$. The other parameters are given in \ref{fig:fig1}.}
\label{fig:fig2}
\end{figure}

\begin{figure}[!ht]
\begin{center}
\includegraphics[width=0.9\columnwidth]{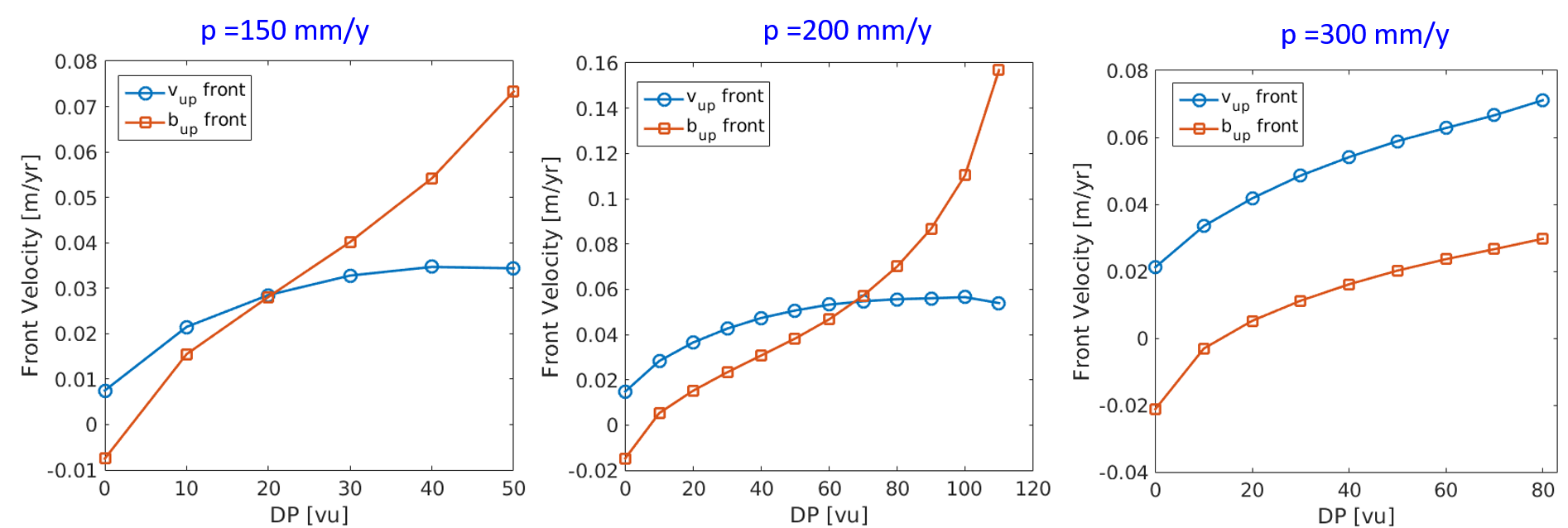}
\end{center}
\caption{Front velocities as a function of DP (drift potential) for (a) $p$=150 mm/yr, (b) $p$=200 mm/yr, and (c) and $p$=300 mm/yr, for the two types of fronts shown in Fig. \ref{fig:fig2}. Note that the velocities are quite small, on the order of a few cm per year. Also note the different axis ranges for the different panels. Generally, the front velocity increases (nonlinearly) with DP.}
\label{fig:fig3}
\end{figure}

\begin{figure}[!ht]
\begin{center}
\includegraphics[width=0.8\columnwidth]{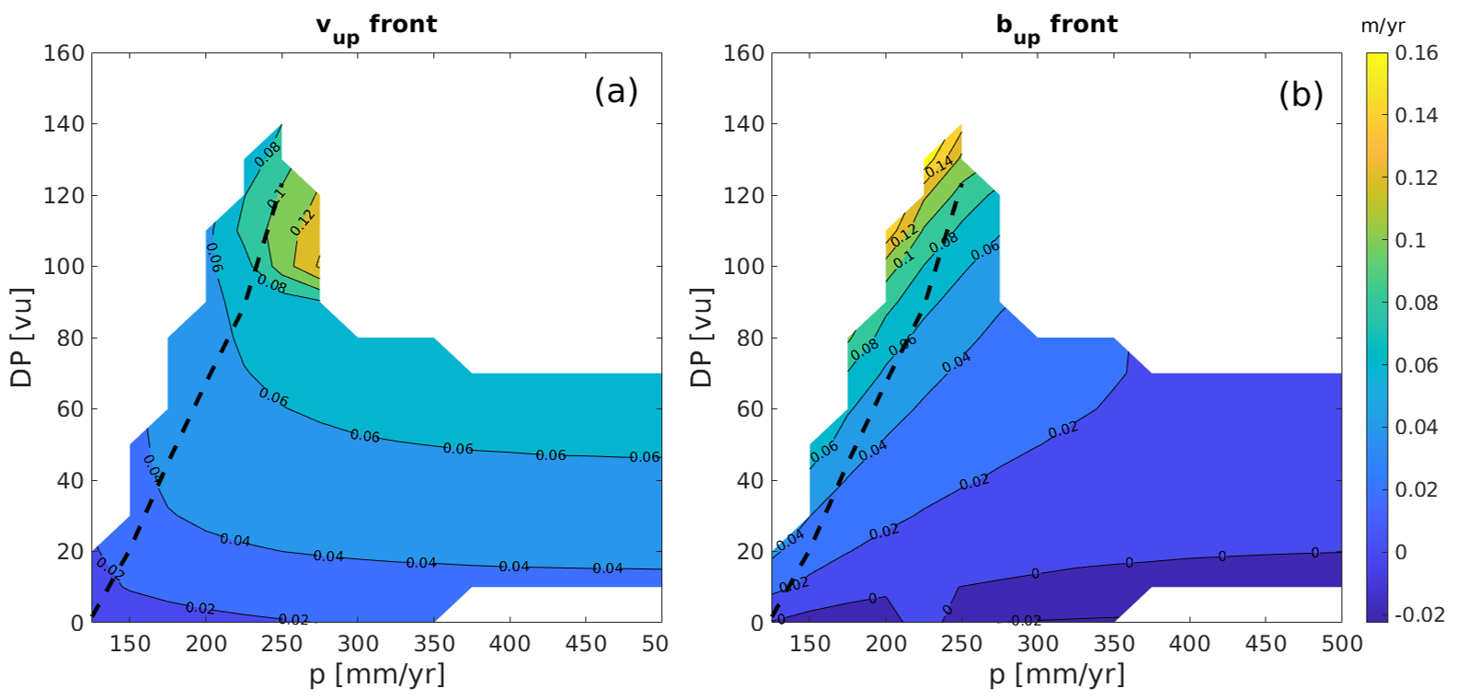}
\end{center}
\caption{Front velocities as a function of $p$ and DP for the (a) $v_{\rm up}$ and (b) $b_{\rm up}$ fronts. The dashed line indicates the location in
the phase space where the front velocities are equal.
The white area indicates the parameter space for which there is only a single stable state. }
\label{fig:fig4}
\end{figure}

Here we study the front solutions that separate the bistable solutions (region ``BS Type I'' in Fig. \ref{fig:fig1}), i.e., one solution is dominated by vegetation and the other by biocrust. Two types of fronts exist: (i) $v_{\rm{up}}$ for which the vegetation cover $v$ is greater than the biocrust cover $b$ in the upwind direction (Fig. \ref{fig:fig2}a) and (ii) $b_{\rm{up}}$ with $b$ greater than $v$ in the upwind direction (Fig. \ref{fig:fig2}b). The dynamics of these two fronts are not symmetrical due to the wind direction that breaks the symmetry in the system. In Fig. \ref{fig:fig3}, we present the front velocity as a function of DP for three precipitation rates. It follows that the fronts are not stationary and propagate with time (order of 1 m per decade) downwind. For sufficiently weak winds (DP=0), the only terms that contribute to the front propagation are the diffusion terms, and in this case, the system is symmetrical, such that flipping the front will yield a front that propagates with the same speed in the opposite direction. The front propagation direction follows the relaxation time to the steady vegetation/biocrust-dominant state; when the relaxation time to the vegetation-dominant state is shorter than the biocrust relaxation time, the front will propagate towards the lower biocrust region. For sufficiently large values of DP and a sufficiently low precipitation rate $p$ (Fig. \ref{fig:fig3}a,b), the velocity of the $b_{\rm{up}}$ front is larger than that of the $v_{\rm{up}}$ front. This is because the domain characterized by high vegetation cover and low biocrust is located downwind; thus, the wind strongly suppresses the vegetation at the front, and consequently, the low vegetation cover domain invades the high vegetation cover. However, as precipitation increases, the effect of this vegetation suppression decreases due to the larger growth rate; this is reflected by the crossing point between the front velocity of the $v_{\rm{up}}$ and $b_{\rm{up}}$ cases, which shifts to higher DP for higher precipitation rates (Fig. \ref{fig:fig3}). In contrast, for the $v_{\rm{up}}$ front, the high biocrust domain is located downwind, and since the biocrust is weakly affected by the wind (it is only affected by sand cover), the front velocity is smaller. Fig. \ref{fig:fig4} summarizes the front velocity as a function of both DP and $p$. It is clear that the front velocity increases as a function of DP. Yet, the situation is more complex for the precipitation rate $p$ for which the front velocity decreases for the $v_{\rm{up}}$ front and increases for the $b_{\rm{up}}$ front.

\begin{figure}[!ht]
\begin{center}
\includegraphics[width=0.8\columnwidth]{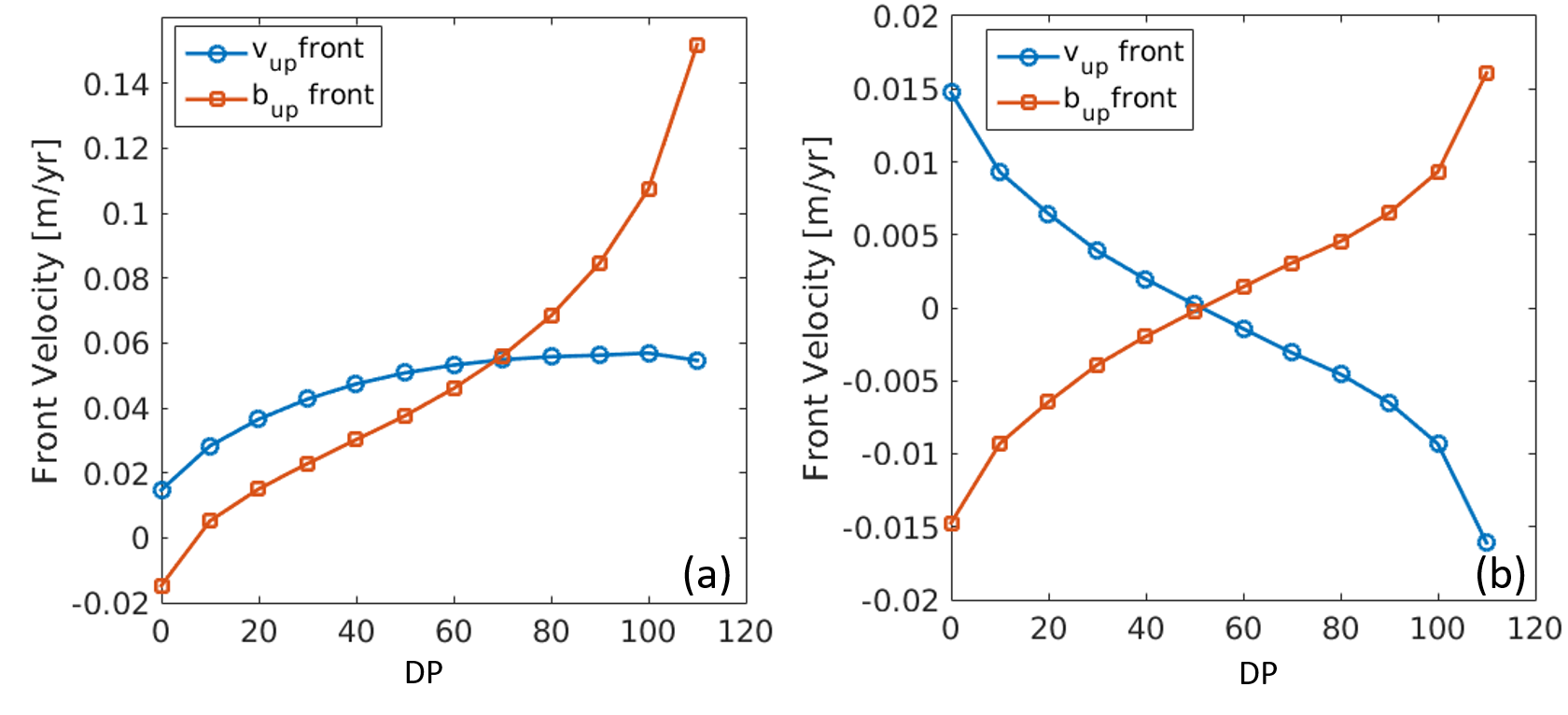}
\end{center}
\caption{Front velocities as a function of DP for $p$=200 mm/yr and for (a) $\beta_v=0$ and for (b) $\kappa=0$. The parameter $\beta_v$ controls the indirect spatial wind effects, and the parameter $\kappa$ controls the
advection of vegetation by the wind, leading to the opposite front progression of $v_{\rm up}$ versus $b_{\rm up}$ when $\kappa=0$ (panel b).}
\label{fig:fig5}
\end{figure}

\begin{figure}[!ht]
\begin{center}
\includegraphics[width=0.5\columnwidth]{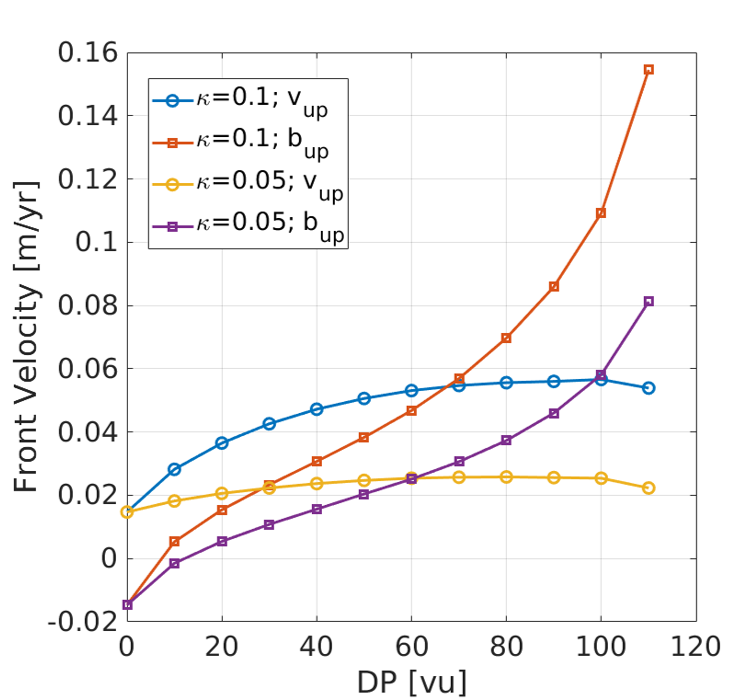}
\end{center}
\caption{Front velocities as a function of DP for $p$=150 mm/yr and for different values of $\kappa$, which is the parameter that controls the vegetation wind advection. The higher the value of $\kappa$, the stronger the wind’s suppression effect on the vegetation.}
\label{fig:fig6}
\end{figure}

Below we study the dependence of the front velocity on the different spatial term parameters. Fig.~\ref{fig:fig5}a shows the front velocity for $\beta_v=0$ as a function of DP, i.e., without the spatial effect of sand transport on vegetation (for $p=200$ mm/yr). The differences between Fig.~\ref{fig:fig5}a and Fig.~\ref{fig:fig3}b (for which $\beta_v\ne 0$) are minor. This is mainly because we used a very small value, $\beta_v= 6.64\times 10^{-5}$, in the simulations in Fig.~\ref{fig:fig3}b, such that when its value is zero, it leads to a negligible effect on the front velocities. In this case, the front velocity of $v_{\rm{up}}$ front is small and converges to an almost constant value for higher $DP$ values.
Fig.~\ref{fig:fig5}b shows the front velocities for $\kappa=0$, i.e., without the wind advection term. In fact, this advection term is the only term that breaks the symmetry in the model, and without it, there is no preference for either direction. This is clearly seen in Fig.~\ref{fig:fig5}b where the front velocity for the $v_{\rm{up}}$ front is a mirror image of the velocity for the $b_{\rm{up}}$ front.  For this case, the dominant term in the equation of $v$ is $- {\beta _v}{\rm{DP}}\left| {\nabla g \cdot \vec k} \right|v$, which is always negative along the front. Its effect is to slow down the $v_{\rm{up}}$ front and to speed up the $b_{\rm{up}}$. The dependence of the front velocities on $\kappa$ as a function of DP is shown in Fig.~\ref{fig:fig6}. $\kappa$ is the parameter that controls the vegetation’s vulnerability to the direct effect of wind at the fronts. Actually, larger $\kappa$ will lead to enhanced advection and to greater front velocity, for both the $v_{\rm{up}}$ and $b_{\rm{up}}$ fronts.

\subsection{Front dynamics in the BS Type II domain}\label{subsec:sec2}
\begin{figure}[!ht]
\begin{center}
\includegraphics[width=0.6\columnwidth]{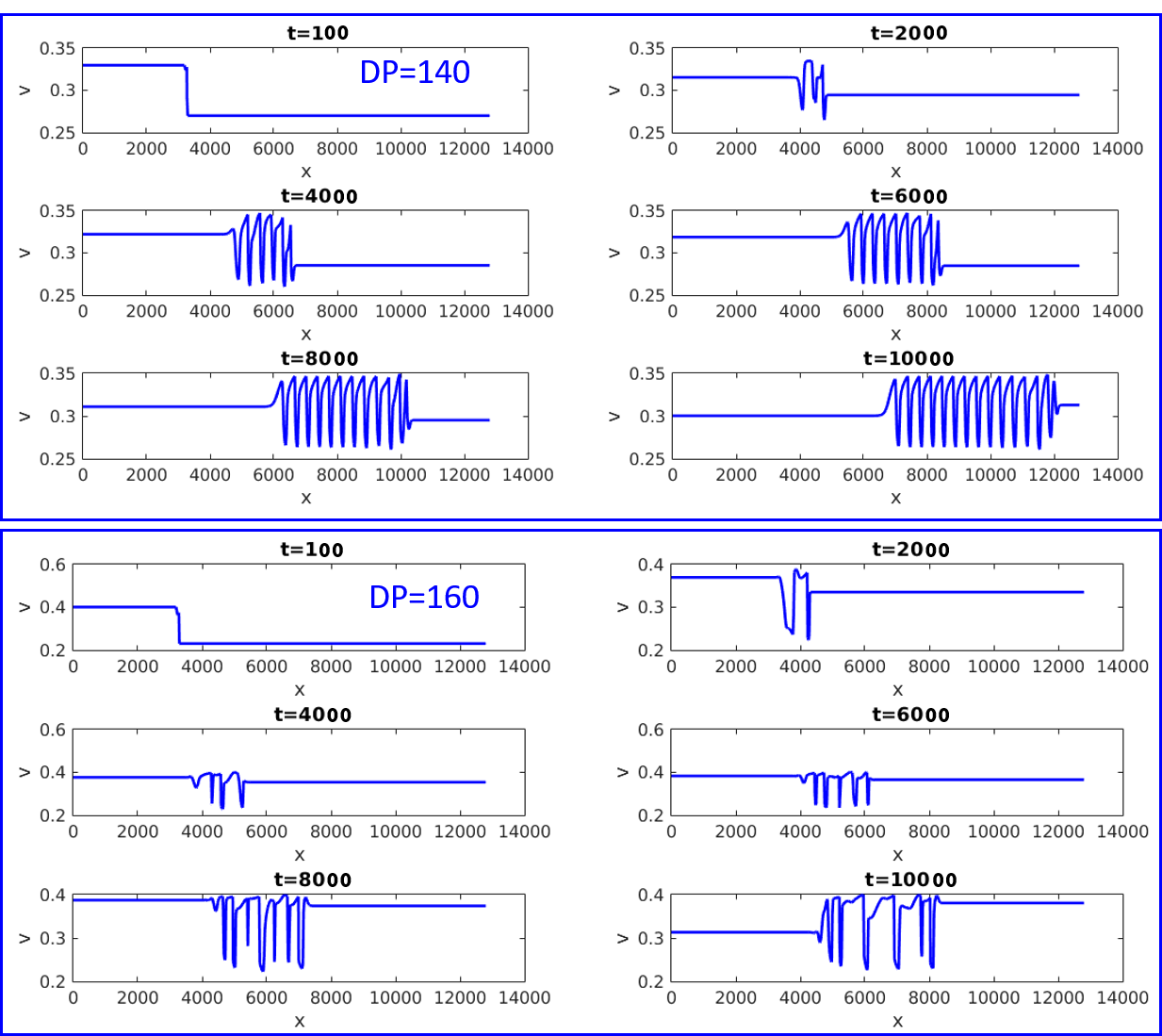}
\end{center}
\caption{Vegetation front dynamics for different times and for DP=140 (upper part) and DP=160 (lower part) for $p$=370 mm/yr. Under this choice of parameters, the spatially uniform regions oscillate with time. As the front propagates to the right (downwind direction), these oscillations produce a sequence of periodic domains that grows in time and extends downwind. The pattern looks ordered with a distinct wavelength for DP=140 but less ordered for DP=160. See Movies M1 and M2 of the Supplementary Material for more details.}
\label{fig:fig7}
\end{figure}

\begin{figure}[!ht]
\begin{center}
\includegraphics[width=0.6\columnwidth]{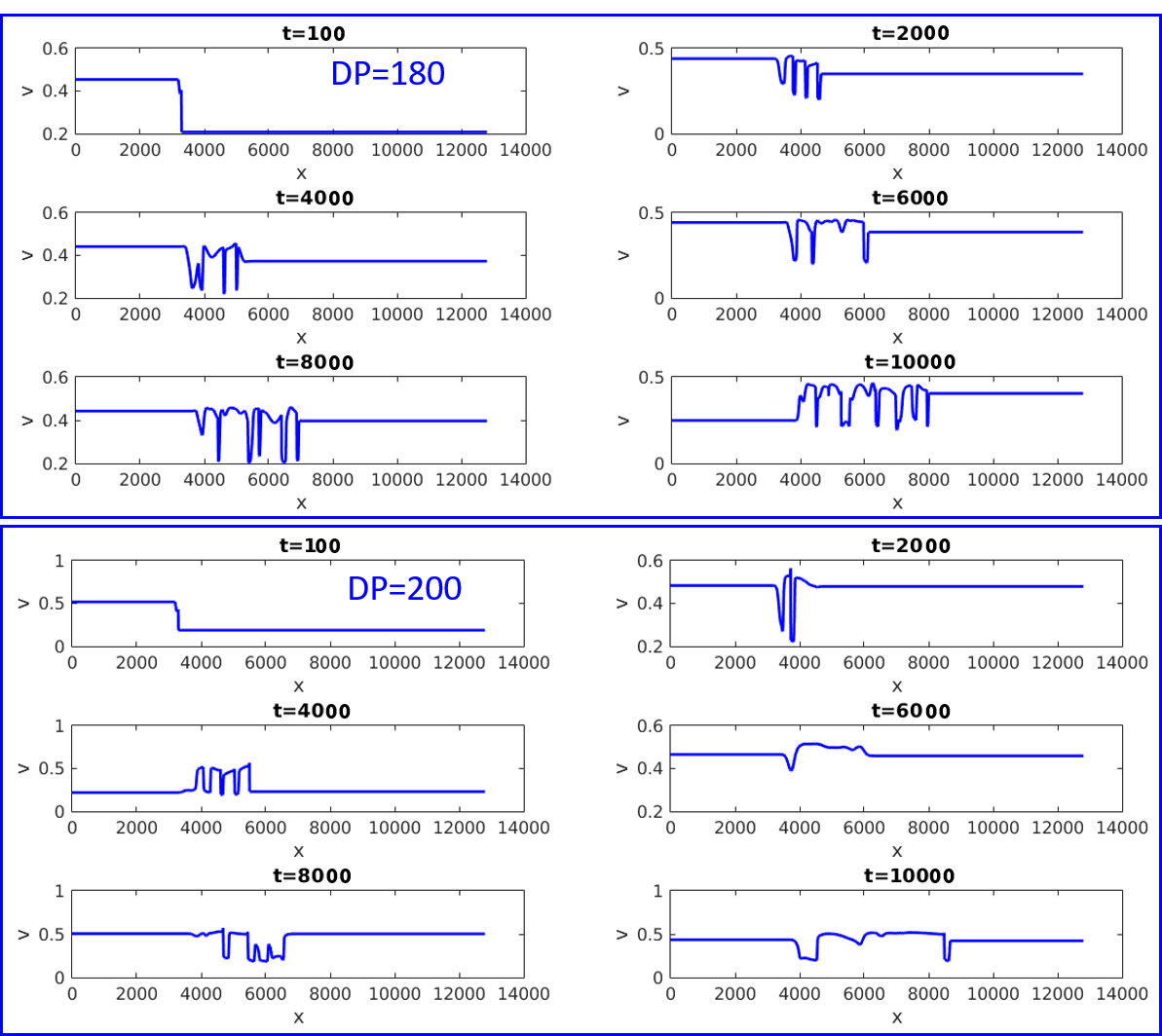}
\end{center}
\caption{Same as Fig.~\ref{fig:fig7} for DP=180 and DP=200. The dynamics is much more disordered. As the front advances downwind, the domains that developed behind it merge together to form wider domains but with different widths. See Movies M3 and M4 of the Supplementary Material for more details.}
\label{fig:fig8}
\end{figure}

The dynamics in the BS Type II domain (Fig. \ref{fig:fig1}) is much more complicated and interesting since in this domain, the states are oscillatory in time, with periods and amplitude that increase with DP (see Fig. 5 in ~\cite{Yizhaq-Ashkenazy-2016:oscillations}). Thus, when the front advances, the states ahead of and behind the leading front oscillate. These oscillations produce a spatially periodic (wave-like) domain that connects the spatially homogeneous regions, as shown for the vegetation cover $v$ in Fig. \ref{fig:fig7} for DP=140 and DP=160. For lower DP values, the periodic pattern is ordered (wavelength and amplitude), and the domains are created at a constant rate (see Movie M1 in the Supplementary Material, which also shows the corresponding biocrust dynamics). For DP=160, the dynamics becomes less ordered where small ``patches'' merge to form wider ``patches.'' For lower DP values, the temporal oscillations are approximately sinusoidal. As DP increases, these become more nonlinear and asymmetrical, with a slowly increasing/decreasing phase that is followed by a rapidly increasing/decreasing phase; these are close to relaxation oscillations. As the oscillation period and asymmetry increase with DP, the pattern becomes less periodic, and the velocity of the spatial pattern is slower. For higher DP values, the pattern becomes even more complex and unpredictable as shown in Fig. \ref{fig:fig8} and Movies M3 and M4 (Supplementary Material) for DP=180 and DP=200. One reason for this complex behavior is that the oscillations created ahead of the leading front become wide enough that small perturbations initiate secondary oscillations inside these domains, such that their sizes are not as constant and periodic as those for the lower DP values (e.g., for DP=140).

\begin{figure}[!ht]
\begin{center}
\includegraphics[width=0.6\columnwidth]{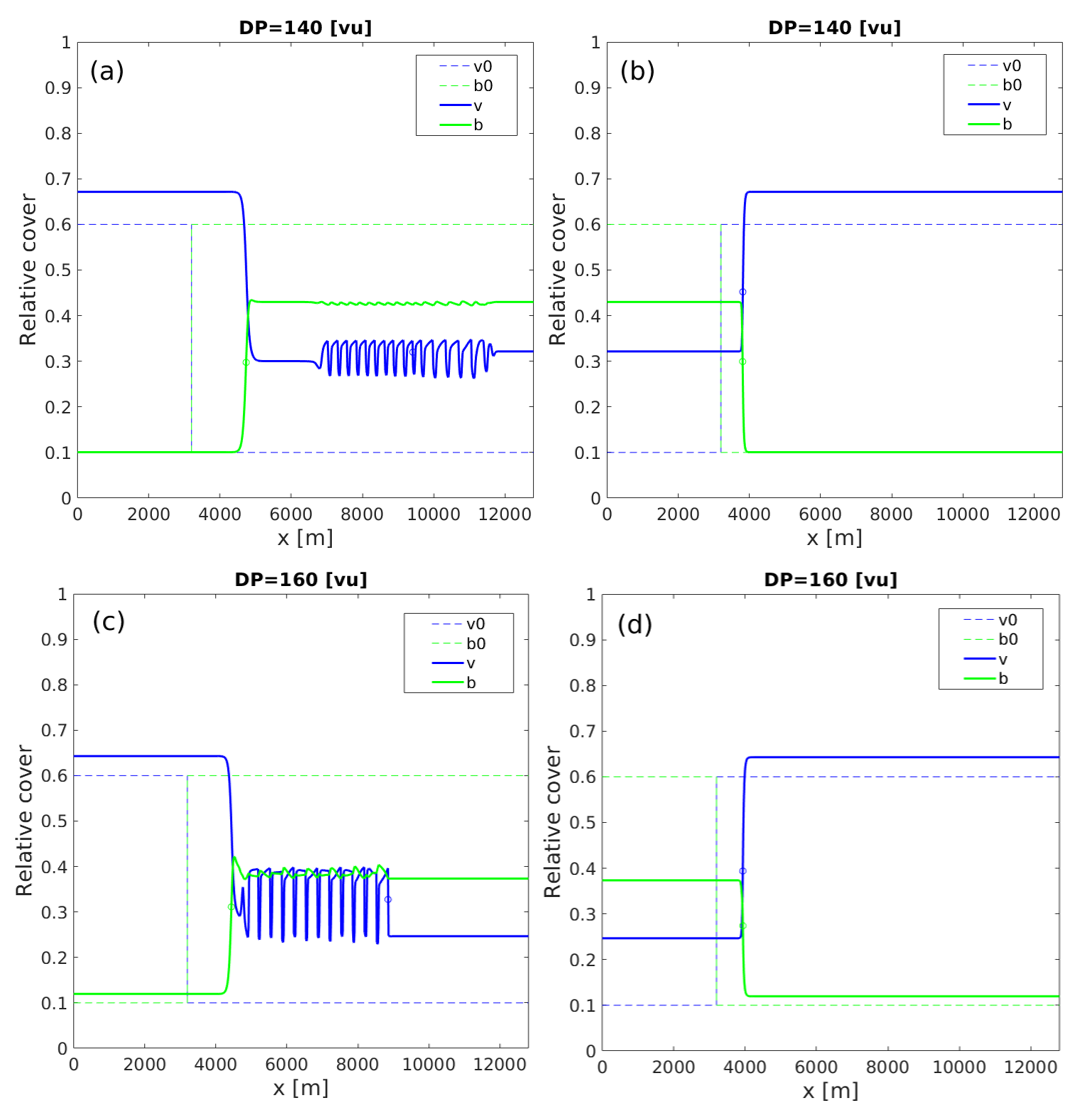}
\end{center}
\caption{The solutions that connect the stable state with the oscillatory state in the BS Type II regime for DP=140 and DP=160. The periodic spatial domains develop when the stationary solution is on the left side of the front (panels a and c). The wind direction is to the right. In the case in which the stationary state is on the right side of the front, the front advances downwind, and no periodic domains form. }
\label{fig:fig9}
\end{figure}

Another type of possible solution in the BS Type II domain is the bistability of a stable state with an oscillatory state; see \cite{Yizhaq-Ashkenazy-2016:oscillations}. In Fig. \ref{fig:fig9}, we initiate one side of the front with the stable solution and the other with the oscillatory solution; the solutions for DP=140 and DP=160 are shown. When the $v_{up}$ state, which is the stationary state, is on the left side of the front and the oscillatory solution is on the right side, the spatial oscillations develop and drift downwind (Fig. \ref{fig:fig9}, a and c; see also movie M5). But when the initial condition is that the oscillatory state is on the left side of the front, the spatial oscillations do not develop, and the front advances downwind (Fig. \ref{fig:fig9}, b and d; see movie M6). Thus, the dynamics is strongly dependent on the initial conditions, and the periodic domains cannot develop and advance into the stationary state; instead, the whole domain on the left side of the front oscillates, and the front advances in a regular manner. The stationary state acts as a barrier that prevents the formation of the small periodic domains. In the opposite case, the stationary state is behind the advancing front, which invades the oscillatory state, and thus, the small domains are free to form.

To quantify the complexity of the evolving front pattern, we propose the measure $L_y$, which is the sum of the absolute values of the difference between two adjacent $v$ (or $b$) values:
\begin{equation}\label{eq:ly}
{L_y} = \sum\limits_{i = 2}^n {\left| {{v_i} - {v_{i - 1}}} \right|} \,.
\end{equation}
$L_y$ defines the total change of $v$ along the variable regions, and it is proportional to its length in the vertical direction (it is zero for a straight line). For regular front propagation, this measure quickly converges to a constant value. Fig. \ref{fig:fig11} shows the evolution of $L_y$ for different values of DP and for slightly different initial solutions. For DP=140 and DP=160, $L_y$ increases monotonically, as the creation of periodic domains behind the leading front is ordered (see Fig. \ref{fig:fig7}), and thus, the curve's length increases over time at an approximately constant rate; still, the periodic creation of new cycles is visible. For DP=180, there is a general increasing trend in $L_y$, but it is more erratic, not monotonic, and can sharply decrease when small domains merge together, causing the length of the curve to sharply decrease. Note that for DP=180, $L_y$ is larger than $L_y$ for DP=140 because the amplitude of the oscillations is larger. For $DP=200$ (the blue curve refers to the solution shown in Fig. \ref{fig:fig8}), there is a very sharp decrease of $L_y$ around $t=6000$ as the solution is almost flat (see Fig. \ref{fig:fig8}). Then, $L_y$ continues to increase till the second sharp decrease at $t=9500$. The less monotonic $L_y$ is, the more erratic the dynamics. The curves, for higher DP values, become more sensitive to the initial conditions. For example, for DP=200, starting with slightly different initial conditions resulted in a different trajectory of $L_y$. This behavior is a sign of chaos in accordance with the complexity and the unpredictability of the solutions for higher DP values. In fact, the curves shown in Fig. \ref{fig:fig11}d are similar to ``random walk'' profiles. The roughness of $L_y$ can be quantified by the $\rho$ defined as:
\begin{equation}\label{eq:rho}
{\rho} = \frac{1}{t} \sum\limits_{i = 2}^n {\left| {{{L_{y,i}}} - {L_{y,i - 1}}} \right|} \,,
\end{equation}
where $t$ is simulation time. Fig. \ref{fig:fig11} shows the average roughness for five different model simulations and their scattering around the average values. $\rho$ increases quadratically with DP and also the deviations from the averages are larger. This is another indication that the solution becomes more complex for larger values of DP.

\section{Discussion and Summary}\label{sec:sec4}

We propose a model for the spatiotemporal dynamics of vegetation and biocrust on sand dunes; the model is based on our previous model of vegetation-biocrust dynamics on sand dunes ~\cite{Yizhaq-Ashkenazy-2016:oscillations} and on a model for the spatiotemporal dynamics of vegetation cover alone on sand dunes \cite{Yizhaq-Ashkenazy-Levin-Tsoar-2013:spatiotemporal}. This model exhibits much more complex and interesting dynamics. We studied the one-dimensional front solution in two DP-$p$ domains where the system shows bistability. In the first domain, the front connects two stationary solutions, while in the second, the front connects two oscillating solutions or connects one oscillatory with one stationary state. In the first case, there are two front types: one that connects the high vegetation cover (and low biocrust cover) with the lower vegetation cover (and high biocrust cover). The second front connects these states in a reserved order. These fronts are not in steady states but advance mainly due to wind advection. Generally, the stronger the wind (quantified by DP), the larger the front velocity; still, the $v_{up}$ front velocity is maximal for DP$\approx$110. The dependence on precipitation is more complicated as the $v_{up}$ front velocity increases with $p$, while the $b_{up}$ front velocity decreases (Fig. \ref{fig:fig4}). This is because wetter conditions favor vegetation growth at the expense of biocrust growth. When $\kappa=0$, i.e., in the absence of wind advection, the fronts can advance in either direction (see Fig. \ref{fig:fig5}). The front velocities are quite small, around 10 cm per year for the parameter set used in the current work. However, the front velocity can become higher for other parameter choices, e.g., larger values of $kappa$. In a previous work regarding the spatiotemporal dynamics of vegetation cover alone without biocrust ~\cite{Yizhaq-Ashkenazy-Levin-Tsoar-2013:spatiotemporal}, the front velocity of the active-fixed dunes was of the order of several meters per year. This much larger front velocity can be attributed, in part, to the different choice of parameters and, in part, to the stabilization effect of the biocrust, which can be eroded only under the action of very strong winds. The equation for the biocrust does not include a wind advection term, such that overall effect of the biocrust is to slow down the front’s velocity.

The fronts in the oscillatory BS Type II regime (Fig. \ref{fig:fig1}) exhibit complex and interesting dynamics. The front connects two oscillatory states. The advance of the front downwind leads to the creation of spatial oscillations. This periodic domain advances downwind. For lower DP values, the spatial oscillations are ordered with a well-defined wavelength and amplitude. The wavelength of the developed oscillatory domain is the result of the front velocity (which is strongly affected by the wind advection) and the oscillation period. The periodic domains are created, in part, due to the oscillations of the rear part of the front and, in part, due to the system’s intrinsic oscillatory nature. The rear part’s movement from the lower state to the upper state creates one domain that moves downwind, as shown in Fig. \ref{fig:fig7}. For higher DP values, the dynamic is very complex and shows a chaotic-like behavior with sensitivity to initial conditions. Fig. \ref{fig:fig8} shows the development of the first domains for DP=180. Here, the oscillation period is longer, and the front itself is unstable. The instability of the front leads to the development of succussive small domains. These domains have their own dynamics; they can merge together in unpredictable ways, and their width is no longer the same. Further work is needed to gain a deeper understanding of this complex dynamic. We also examined the case where one side of the initial front is in a stationary state, while the other is in an oscillatory state. In this case, it is possible to obtain spatial oscillations or a simple front progression with spatial oscillation. This depends on the wind direction as shown in Fig. \ref{fig:fig9}. The oscillations develop when the stationary state is upwind of the oscillatory state ($v_{up}$ front, Fig. \ref{fig:fig9} a and c), as in this case, the stationary state invades the oscillatory state. Since the front always advances downwind, in the opposite case, i.e., when the oscillatory state invades the stationary state ($b_{up}$ front, Fig. \ref{fig:fig9} b and d), the front advances, and no periodic domains form. In this case, the stationary state blocks the formation of the periodic domain, as the front is pinned to the stationary solution.

The aim of the current work was to develop a spatial model of vegetation and biocrust cover on sand dunes; we show that this model has very interesting and complex dynamics. In the real world, there are many factors that can hinder the observations of such long-term spatial oscillations. The forcing of the climate conditions, such as wind and precipitation, is not uniform either in space or time, as in the model, but is instead intermittent; moreover, the wind is not always unidirectional. Spatial heterogeneities in the other parameters of the model can alter these oscillations. In addition, the model ignores the feedback between dunes and climate, for example, due to the albedo effect of sand dunes ~\cite{Ashkenazy-Shilo-2018:sand}. As was shown in ~\cite{Yizhaq-Ashkenazy-2016:oscillations}, fluctuations in precipitation can eliminate the model's oscillations and shift the system to one of the bistable states and thus to the front dynamics in BS Type I. This result suggests that it would be difficult to observe these predicted complex spatial oscillations in reality due to their long period and the interaction with the external variability of the climate system. Yet, the oscillations and the complex behavior on a time scale of hundreds of years that we observe in the model can be associated with the red-noise spectrum of paleoclimate records; see, e.g., \cite{Hasselmann-1976:stochastic, Ashkenazy-Baker-Gildor-et-al-2003:nonlinearity}. A further study is needed to gain a deeper understanding of the behavior of the dynamics in the oscillatory regime.

\section*{Acknowledgments}
The researcher leading this work received funding from the Israel Science Foundation (ISF), grant number 156/16. We thank Golan Bel for helpful discussions.

\section*{Data availability statement}
All data that support the findings of this study are included within the article (and any supplementary files).

\section*{References}

\providecommand{\newblock}{}

\newpage
\section*{Figure Legends}

\begin{figure}[!ht]
\begin{center}
\includegraphics[width=0.5\columnwidth]{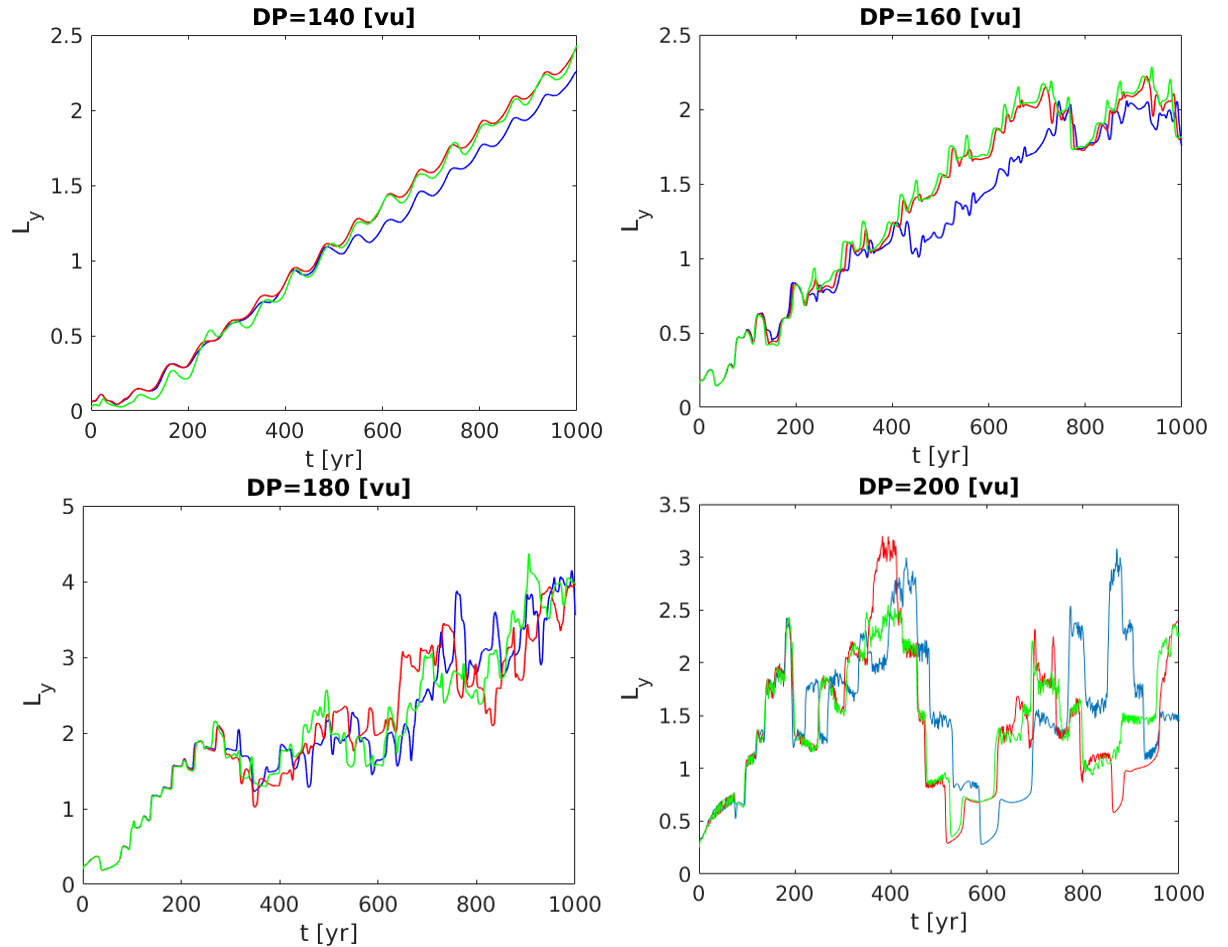}
\end{center}
\caption{$L_y$ (Eq. \ref{eq:ly}) as a function of time for different values of DP. $L_y$ is proportional to the length of the front solution. For lower DP, $L_y$ increases monotonically and becomes more erratic for higher DP values. For each DP value, the results of three slightly different initial conditions are shown. The sensitivity of $L_y$ to the initial conditions increases with DP, indicating that the solution becomes more complex. }
\label{fig:fig11}
\end{figure}

\begin{figure}[!ht]
\begin{center}
\includegraphics[width=0.5\columnwidth]{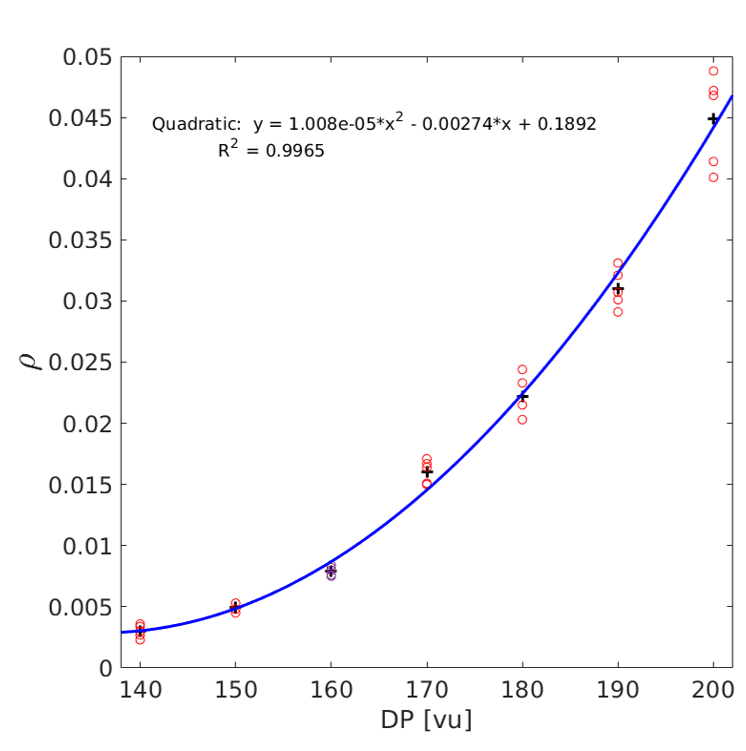}
\end{center}
\caption{The roughness $\rho$ (Eq. \ref{eq:rho}) as a function of DP. The plus signs indicate the average of five realizations of the model with slightly different initial conditions, and the red circles show the value of each one of the realizations.  The roughness increases quadratically with DP. Note that the scattering around the average increases with DP, which is another indication that the solution is more complex for larger values of DP.}
\label{fig:fig11}
\end{figure}



\end{document}